\documentclass[twocolumn,pra,aps,showpacs,amsmath,amssymb]{revtex4}
\usepackage{graphicx}
\usepackage{dcolumn}
\usepackage{bm}

\begin{document}
\title{One-way propagation of light in Born-Infeld-like metamaterials}
\author{Vitorio A. De Lorenci$\mbox{}^{1,2}$}
\email{delorenci@unifei.edu.br}
\author{Jonas P. Pereira$\mbox{}^{3,4,5}$}
 \email{jonaspedro.pereira@icranet.org}

 \affiliation{$\mbox{}^{1}$Instituto de F\'isica e Qu\'imica,
Universidade Federal de Itajub\'a, Itajub\'a, MG 37500-903, Brazil}
\affiliation{$\mbox{}^{2}$Institute of Cosmology, Department of Physics and Astronomy,
Tufts University, Medford, MA 02155, USA }
\affiliation{$\mbox{}^{3}$Universit\'e de Nice Sophia Antipolis, 28 Av. de Valrose, 06103 Nice Cedex 2, France}
\affiliation{$\mbox{}^{4}$Dipartimento di Fisica and ICRA, Universit\`{a} di
Roma - La Sapienza, P.le Aldo Moro 5, I-00185 Rome, Italy}
\affiliation{$\mbox{}^{5}$ ICRANet, Piazza della Repubblica 10, I-65122 Pescara, Italy}

\date{\today}

\begin{abstract}
We propose and investigate a family of nonlinear metamaterials in which light rays propagate just in one direction -- one-way propagation. Furthermore, we argue how such nonlinear media could provide an analog model for investigating the Born-Infeld Lagrangian in the
realm of fields larger than its scale field.
\end{abstract}

\maketitle

\section{Introduction}
In order to circumvent the point-charge singularity appearing in the Maxwell theory, alternative Lagrangians to the electromagnetism have been proposed or emerged (as effective theories) by other more fundamental theories. The common property of these Lagrangians is that they are built up just in terms of the two local Lorentz invariants of the electromagnetism \cite{LandauEM1987,Jackson1998} in a nonlinear way.
We cite for example the Born-Infeld Lagrangian \cite{BILag}. Under the theoretical point of view, such a Lagrangian is interesting in the sense it renders a finite self-energy to the electron \cite{BILag}, it leads to an exact solution in general relativity \cite{Garcia1984, Nora2001}, and a similar form to it appears as an effective Lagrangian to the low-energy limit in string theory \cite{Rasheed2008, Labun2010}. Another important example is the one loop effective Euler--Heisenberg Lagrangian \cite{Heisenberg1936,Schwinger1951}. This Lagrangian has been applied, for instance, in the astrophysical scenario by attempting to enhance the description of pulsars \cite{Dupays2008}, black holes \cite{DeLorenci2001}, gravitational lensing \cite{Herman2006}, etc. For a review on these effective Lagrangians, see Ref.~\cite{Ruffini2010} and references therein.

For any Lagrangian, the effective medium interpretation can be applied \cite{BILag,Ruffini2010}. The converse is not necessarily true,
as the dielectric tensors must satisfy certain differentiability constraints in order to assure the existence of a Lagrangian. It would be true, for instance, for a medium whose permittivity and permeability are just dependent upon the electric and magnetic fields, respectively. Hence the aforesaid map could be useful for testing analogously nonlinear theories proposed in the literature (e.g. Refs.~\cite{Garcia1998, Garcia1999, novello2004}) by means of engineering adequate metamaterials \cite{metamaterials}, as in general they are unlikely to be found in nature. The rudiments of this new area of research came from Veselago \cite{veselago}, with his studies on media endowed with negative dielectric coefficients. Metamaterials display very unusual effects \cite{smith2000,   Landy2008, smolyaninov2009, schurig2006, pendry2000}, which can lead to potential applications, as well as to analogous tests of miscellaneous areas of physics \cite{smolyaninov2010}. Concerning nonlinear metamaterials, advances are also on their way \cite{rose2011}. Trirefringence \cite{delorenci2012}, for instance, would be an effect such media may display. For further effects, see Ref.~\cite{rose2011} and references therein.

By one-way propagation, we signify a complete asymmetry of light propagation such that if light propagates forwards (in an arbitrary direction) it can not propagate backwards.
It requires breaking time-reversal and parity symmetries of the system \cite{figotin2001, cheng2011}.
It is possible to tailor some metamaterials to present asymmetry in light propagation \cite{menzel2010}. Conditions have already been reached to obtain one-way propagation in terms of both phase and group velocities in some photonic crystals \cite{cheng2011}. In this paper we address the issue of one-way propagation of light in another context, namely, the nonlinear media described by dielectric coefficients, by proposing a model where such an effect is expected to take place. For doing so, we rely on the method of field disturbances applied to nonlinear media in the limit of geometric optics \cite{LandauEM1987,born, LandauCM1984}. For a limiting case, we show that the proposed medium can be mapped onto the Born-Infeld Lagrangian such that the one-way effect can be used to assess such theory. Thus, this investigation would also be useful as an analog model, since the Born-Infeld scale field (as large as $10^{15} esu\simeq 10^{19}V/m$) is unlikely to be reached on terrestrial laboratories at the present time.

In the next section we elaborate on the general problem of propagation of disturbances in the limit of geometric optics by making use of the formalism developed by Hadamard and Papapetrou \cite{Hadamard}. Section \ref{one-way medium} is devoted to a thorough study of a family of nonlinear media exhibiting one-way propagation in terms of rays, whose speed and direction are governed
by the group velocity \cite{LandauEM1987}, in the form of birefringence. We close this paper with section \ref{discussion}, where some points will be elaborated. We work in Cartesian coordinates. Units are set such that $c=1$.

\section{General description of disturbances in material media }
\label{disturbances}
Maxwell's equations in material media with sources can be summarized as
\begin{eqnarray}
\sum_{i=1}^3\partial_iD_i&=&\rho,\;\;\;\;\;\sum_{j,k=1}^3\epsilon_{ijk}\partial_jE_k=-\partial_tB_i,\label{1a}\\
\sum_{i=1}^3\partial_iB_i&=&0,\;\;\;\;\;\sum_{k,l=1}^3\epsilon_{ikl}\partial_kH_l=\partial_tD_i+ j_i\label{01},
\end{eqnarray}
where $D_i$ and $H_i$ are the i{\it th} components of the induced fields,
whilst $E_j$ and $B_j$ are the j{\it th} components of the strength ones, and $\rho$ and $j_i$ are the free charge density and the i{\it th} free current component density in the medium, respectively. Besides, the totally antisymmetric quantity $\epsilon_{ijk}$ was defined \cite{LandauEM1987} such that $\epsilon_{123}=+1$, $\partial_i\doteq {\partial}/{\partial x_i}$, and $\partial_t\doteq {\partial}/{\partial t}$.
For making the equations of electrodynamics in material media complete, the constitutive relations
\begin{equation}
D_i = \sum_{j=1}^3\varepsilon_{ij}(\vec{E},\vec{B})E_j,\;\;\;
H_i = \sum_{j=1}^3\mu^{\scriptscriptstyle{ -1}}_{ij}(\vec{E},\vec{B})B_j\label{02},
\end{equation}
must be assumed; we call $\varepsilon_{ij}$ the permittivity tensor of the material medium and $\mu^{\scriptscriptstyle{ -1}}_{ij}$ its inverse permeability tensor. For the vacuum, for instance, the aforementioned dielectric tensors are written as $\varepsilon_{ij} = \varepsilon_0 \delta_{ij}$ and $\mu^{\scriptscriptstyle{ -1}}_{ij} = (1/\mu_0)\delta_{ij}$, $\delta_{ij}$ such that it is $1$ iff $i=j$, otherwise it is $0$. These tensors encompass all the electromagnetic properties of a material medium and in general are dependent upon the strength fields.

For investigating the propagation of electromagnetic waves in material media in the limit of geometric optics, the method of field disturbances \cite{Hadamard} can be used. This can be succinctly enunciated as follows. Assume an at least ${\cal C}^2$ hypersurface $\Psi(t,\vec{x})=0$, named $\Sigma$, that splits the spacetime into two disjointed regions. These regions are formed by the space-time points $P^-$ such that $\Psi <0$ and $P^+$ where $\Psi >0$.
The discontinuity of an arbitrary function dependent upon the space-time coordinates at an arbitrary point $P$ belonging to $\Sigma$
is defined as
\begin{equation}
\label{discontinuity}
\left[f(P)\right]_{\Sigma}
\doteq \lim_{\epsilon\rightarrow 0^+}
\left[f(P+\epsilon) - f(P-\epsilon)\right].
\end{equation}
Assuming that the electromagnetic fields are continuous on $\Sigma$, and that
the latter is the eikonal of the disturbances under interest, following Hadamard and Papapetrou \cite{Hadamard}, the first derivatives of the resultant fields are not
continuous on $\Sigma$ and behave as \cite{delorenci2012}
\begin{eqnarray}
{\left[\partial_{t}E_i\right]}_{\Sigma}&\!\!\!=\!\!&\omega e_i\,,\;\;\;\;\;\;
{\left[\partial_{t}B_i\right]}_{\Sigma}=\omega b_i\,,
\label{q1}\\
{\left[\partial_i E_j \right]}_{\Sigma}&\!=\!&-q_i e_j,\;\;\;
{\left[\partial_i B_j \right]}_{\Sigma}\,\!=\!-q_i b_j,
\label{q2}
\end{eqnarray}
where $e_j$ and $\beta_j$ are related to the derivatives of the electric and magnetic
fields on $\Sigma$ and correspond to the j{\it th} components of the electric and magnetic polarization vectors of the propagating
waves \cite{delorenci2012}, $\vec{e}$ and $\vec{\beta}$, respectively. The angular frequency and
the i{\it th} component of the wave vector are defined by $\omega$ and $q_i$, respectively. Besides, in the present case of $\Sigma$ as the eikonal of the disturbances, its orthogonal four-vector is the wave-four-vector $k_{\mu}=(\omega, -\vec{q})$. Physically, one can understand Eqs.~(\ref{q1}) and (\ref{q2}) as the possibility of having plane waves in a small region around any point on the eikonal in the limit of geometric optics and of a linearization
of the field equations in this limit. Furthermore, since the eikonal was assumed to be an equipotential and well defined hypersurface, the field discontinuities must be just in a direction perpendicular to it, as evidenced by the frequency and components of the wave vector in Eqs.~(\ref{q1}) and (\ref{q2}).

By substituting Eqs.~(\ref{q1}) and (\ref{q2}) in Eqs.~(\ref{1a}) and (\ref{01}) and assuming that free charge densities and currents have a zero discontinuity across $\Sigma$, after some
simplifications, one has the following equations for the polarization vectors $\vec{\beta}$ and $\vec{e}$:
\begin{equation}
\vec{\beta}=\frac{1}{\omega}(\vec{q}\times \vec{e})\label{011}
\end{equation}
and
\begin{equation}
\sum_{j=1}^3Z_{ij}e_j=0 \label{05},
\end{equation}
where
\begin{eqnarray}
Z_{ij}&=&\frac{|\vec{q}|}{\omega}\sum_{k,l,m=1}^3\left(\frac{\partial \varepsilon_{ik}}{\partial B_l}\epsilon_{lmj}E_k + \frac{\partial \mu^{\scriptscriptstyle{ -1}}_{lk}}{\partial E_j}\epsilon_{iml}B_k\right)\hat{q}_m +
\nonumber \\
&&+ \frac{|\vec{q}|^2}{\omega^2}\sum_{k,l,m,p=1}^3 \epsilon_{ilm}\epsilon_{pkj}H_{mp}\hat{q}_{l}\hat{q}_k + C_{ij},
\label{06}
\end{eqnarray}
with
\begin{equation}
C_{ij}\doteq\varepsilon_{ij}+\sum_{k=1}^3\frac{\partial \varepsilon_{ik}}{\partial E_j}E_k\label{08},
\end{equation}
\begin{equation}
H_{ij}\doteq \mu^{\scriptscriptstyle{ -1}}_{ij}+\sum_{k=1}^3\frac{\partial \mu^{\scriptscriptstyle{ -1}}_{ik}}{\partial B_j}B_k\label{09}.
\end{equation}
In addition, we have defined $X^2=\sum_{i=1}^3 X_iX_i$, is the square modulus of the field $\vec{X}$ with Cartesian components $X_i$, and the j{\it th} component of the unit wave vector $\hat q$ is defined as $\hat{q}_j\doteq q_j/|\vec{q}|$. For a tensorial description of the propagation of disturbances in material media, see for instance Refs.~\cite{delorenci2008,delorenci2002}.

In order to have nontrivial solutions to Eq.~(\ref{06}) concerning the electric polarization, one has to impose
$\det( Z_{ij})\,=0$, that is \cite{silva1998}
\begin{eqnarray}
({Z_{1}})^{3}\,
-\,3{Z_{1}}{Z_{2}}\,+2\,{Z_{3}}=0,
\label{012}
\end{eqnarray}
where we defined,
\begin{eqnarray}
Z_{1} &\doteq& \sum_{i=1}^{3} Z_{ii},
\label{tracesa}
\\
Z_{2} &\doteq& \sum_{i,j=1}^{3} Z_{ij}\,Z_{ji},
\label{tracesb}
\\
Z_{3} &\doteq& \sum_{i,j,l=1}^{3}Z_{ij}\,Z_{jl}\,Z_{li}.
\label{tracesc}
\end{eqnarray}
Equation (\ref{012}) is called Fresnel's equation and it gives the dispersion relation of the medium under interest.

\section{A model for media presenting one-way propagation of light}
\label{one-way medium}

\begin{figure}[!hbt]
\leavevmode
\centering
\includegraphics[scale = .65]{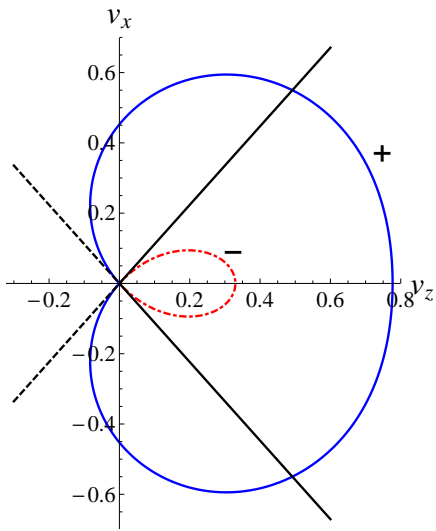}
\caption{{\small\sf (Color online) Normal surfaces \cite{LandauEM1987,born} for the media with dielectric coefficients given by Eqs.~(\ref{03}) and (\ref{04}) for $E=1.5\,b$ $B=1.2\,b$ and $\alpha=0.5$. Such surfaces are symmetric about the $z$ axis, as clearly seen by Eq.~(\ref{16}). The aforementioned media do not generate ordinary solutions to the Fresnel equation but two extraordinary ones, denoted by ``$+$'' and ``$-$'' and depicted by the thick and dot-dashed curves, respectively. If $E>b$, then a region will always exist where the birefringence effect takes place, as clearly shown by Eq.~(\ref{17}) and for the particular example, encompassed by the two thick straight lines. This region increases when one increases the value of the controllable electric field, as expressed again by Eq.~(\ref{17}). In the complementary region, one refraction and no wave propagation are also present, as depicted between the thick and dashed straight lines and between the two dashed straight lines, respectively.}}\label{fig1}
\end{figure}

In this work, we shall be interested in the propagation of weak electromagnetic disturbances in symmetric media described by
\begin{equation}
\varepsilon_{ij}=\frac{\varepsilon}{\sqrt{1+\frac{B^2-E^2}{b^2}-\frac{(\vec{E}\cdot\vec{B})^2}{b^4}}}\left[\delta_{ij}+\frac{B_iB_j}{b^2} \right]\label{03}
\end{equation}
and
\begin{equation}
\mu^{\scriptscriptstyle{ -1}}_{ij}=\frac{1}{\mu}\frac{1}{\sqrt{1+\frac{B^2-E^2}{b^2}-\frac{(\vec{E}\cdot\vec{B})^2}{b^4}}}\left[\delta_{ij}-\frac{E_iE_j}{b^2} \right]\label{04},
\end{equation}
with $b$ a parameter characterizing each medium, and $\varepsilon$ and $\mu$ its (isotropic) permittivity and permeability in the absence of fields. The motivations for the investigation of media described by Eqs.~(\ref{03}) and (\ref{04}) will be given
later.
As it is reasonable, we shall consider hereafter that the fields of the waves are much smaller than their controllable counterparts, $\vec{E}_c$ and $\vec{B}_c$, obtained either by means of prescribed sources in the material media or by convenient pumping fields.
Hence, in Eqs.~(\ref{05})--(\ref{09}), one should rightfully assume $\vec{E}\approx \vec{E}_{c}$ and $\vec{B}\approx \vec{B}_{c}$.
In order to simplify our reasoning, evidencing the physical nature of the effect under interest, we shall assume that the controllable fields are
constant and are written as $\vec{E}_{c}\doteq E\hat{x}$ and $\vec{B}_{c}\doteq B\hat{y}$. Besides, we shall assume that the wave vector can just lie in
the $xz$ plane defined by the above coordinate system. Hence, the unit wave vector can be decomposed as $\hat{q}_x=\sin\theta$ and $\hat{q}_z=\cos\theta$, where $\theta$ is the angle $\vec{q}$ makes with the $z$ axis. For these configurations, it can be easily seen from Eqs.~(\ref{03}) and (\ref{04}) that the
dielectric tensors are diagonal. The same happens with the quantities $C_{ij}$ and $H_{ij}$, as defined by Eqs. (\ref{08}) and (\ref{09}), respectively. Subtleties are present just concerning the first term on the right hand side of Eq.~(\ref{06}).
Specialized to the aforementioned conditions, trivial but tedious calculations for Eqs.~(\ref{tracesa})--(\ref{tracesc}), taking into account the dielectric coefficients
given by Eqs.~(\ref{03}) and (\ref{04}), allows one to cast the Fresnel equation, Eq.~(\ref{012}), as
\begin{equation}
(B\omega-Eq\cos\theta)(B\omega-E\alpha q\cos\theta)-b^2(\alpha q^2-\omega^2)=0
\label{014},
\end{equation}
where $\alpha \doteq 1/(\mu\varepsilon)$ and $q\doteq |\vec{q}|$.
The solution to the above equation can be presented as
\begin{equation}
v_{\pm} = \frac{EB(1+\alpha)\cos\theta}{2(b^2+B^2)} \pm b\sqrt{\frac{\alpha(\sin^2\theta+{\cal J}\cos^2\theta)}{b^2+B^2}}\label{16},
\end{equation}
where we have defined the phase velocity as $v^2\doteq\omega^2/q^2$ and
\begin{equation}
{\cal J}\doteq 1-\frac{E^2}{b^2+B^2}+\frac{E^2B^2(1-\alpha)^2}{4b^2\alpha(b^2+B^2)}\label{24}.
\end{equation}
For the case $b\rightarrow \infty$, from Eq.~(\ref{16}), the phase velocities tend to $\pm \sqrt{\alpha}$, as we already expected from the definition of $\alpha$. One also sees from the same equation that the family of media under interest is such that no ordinary waves (isotropic waves \cite{LandauEM1987, born}) propagate.
The above equations show us that the birefringence effect (defined here as the presence of two waves in a same wave vector direction, which implies that upon refraction two rays will propagate in the medium under interest; see Eqs.~(\ref{18})--(\ref{20}) and Refs.~\cite{LandauEM1987,delorenci2012,born}) will take place only in the region of the
$xz$ plane defined by
\begin{equation}
-\arccos\left(\frac{b}{E}\right)< \theta < \arccos\left(\frac{b}{E}\right)\label{17}.
\end{equation}
Hence, this region will exist iff $E>b$. From now on, we shall consider this to be the case. It implies that the underlying medium (picking out a particular $b$, $\varepsilon$ and $\mu$) must present some negative dielectric tensor components, since in principle $\mu$ and $\varepsilon$ could also be negative, for allowing the birefringence effect. Notice that no condition is imposed on the controllable magnetic field for having the aforesaid optical effect. This will be in sharp contrast with the group velocity analysis, as we shall show later.
Once the limit of geometric optics is just meaningful for media where losses are negligible \cite{LandauCM1984}, we will assume that
\begin{equation}
b^2+B^2-E^2>0\label{17b}.
\end{equation}
A case where the controllable fields lead to the phenomenon of birefringence in depicted in Fig.~\ref{fig1}. For the selected set of the fields, the birefringent region is encompassed by the two thick straight lines.
For the angles limited by the two dashed straight lines, no waves propagate. For the remaining angles, just an extraordinary wave is present.
Since the deduction of Eq.~(\ref{014}) has been done in the limit of geometric optics, it follows that the physically relevant velocities to be
analyzed are the group velocities, that define the speed of the rays \cite{LandauEM1987}. This is the case once packets could be formed due to the linearization process brought by this limit when weak disturbances are present \cite{birula1970}. The group velocities can be easily obtained from an implicit differentiation of the dispersion relation [Eq.~(\ref{014})] \cite{smith2003}
by considering that $\omega = \omega (\vec{q})$ and $\vec{q}= q\hat{q}$. From fields and waves satisfying our
previous conditions, it is easy to show that the extraordinary group velocities associated with the extraordinary phase velocities, here denoted by $\vec{u}$,
are given by
\begin{equation}
\vec{u}\doteq\frac{\partial\omega}{\partial \vec{q}}=u_x\hat{x} + u_z\hat{z}\label{18},
\end{equation}
where
\begin{equation}
u_x=\frac{2b^2\alpha\sin\theta}{2(B^2+b^2)v-EB(1+\alpha)\cos\theta}\label{19}
\end{equation}
and
\begin{equation}
u_z=\frac{BE(1+\alpha)v -2\alpha(E^2-b^2)\cos\theta}{2(B^2+b^2)v-EB(1+\alpha)\cos\theta}\label{20},
\end{equation}
where $v$ is a shortcut to $v_{\pm}$, given by Eq.~(\ref{16}).
Hence, the extraordinary group velocities remain in the same plane as the phase velocities.
It is possible to obtain analytically the main features concerning the group velocities. If one defines  $\varphi$ as the angle between the group
velocity and the $z$ axis, then from Eqs.~(\ref{19}) and (\ref{20}),
\begin{equation}
\tan\varphi= \frac{u_x}{u_z}=\frac{2b^2\alpha\sin\theta}{BE(1+\alpha)v -2\alpha(E^2-b^2)\cos\theta}\label{21}.
\end{equation}
One sees that when $\theta \longrightarrow -\theta$, $\varphi \longrightarrow -\varphi$. Then, it follows from Eqs.~(\ref{19}) and (\ref{20}) that $|\vec{u}|$ is symmetric about the $z$ axis, or $|\vec{u}(\varphi)| = |\vec{u}(-\varphi)|$. Hence it suffices just analyzing the region  $0\leq \varphi \leq \pi$.

\begin{figure}[!hbt]
\leavevmode
\centering
\includegraphics[scale = .60]{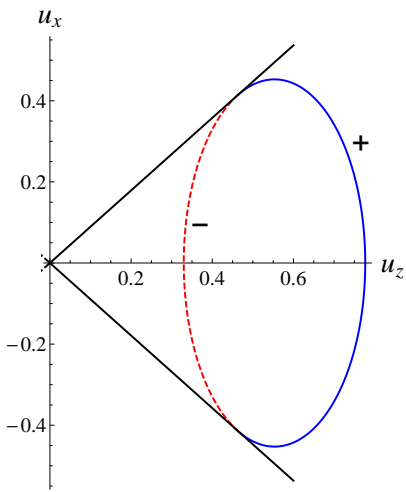}
\caption{{\small\sf (Color online) Ray surfaces \cite{LandauEM1987,born} for the same medium and parameters as in Fig.~\ref{fig1}. As the phase velocities, the group velocities solutions are also symmetric about the $z$ axis. Whenever $E>b$ and Eq.~(\ref{17b}) are valid, just a birefringent region exists, as given by Eq.~(\ref{26}). For the previously chosen parameters, such a birefringent region is encompassed by the two thick straight lines.  Inside this region, for any direction chosen, just two extraordinary rays are present (related to two different extraordinary waves), denoted by ``$+$'' and ``$-$'' and depicted by the thick and dashed curves, respectively; for the set of parameters chosen they are depicted by the thick and dashed curves, respectively. Outside the birefringent region, rays cannot be found. The size of this region decreases with the increase of the electric field.}}\label{fig2}
\end{figure}

If one substitutes Eq.~(\ref{16}) into Eq.~(\ref{21}), one obtains
\begin{eqnarray}
({\cal V}\,\tan^2\varphi-1)\,\tan^2\theta +2{\cal J}\,\tan\varphi\, \tan\theta +\nonumber\\
+{\cal J}({\cal V}-{\cal J})\,\tan^2\varphi &=& 0\label{22},
\end{eqnarray}
where
\begin{equation}
{\cal V}\doteq \frac{E^2B^2(1+\alpha)^2}{4\alpha b^2(b^2+B^2)}\label{23}.
\end{equation}
The solution to Eq.~(\ref{22}) is
\begin{equation}
\tan\theta =\frac{{\cal J}\tan\varphi}{1-{\cal V}\tan^2\varphi}\left[1\pm \sqrt{\frac{{\cal V}}{{\cal J}} \left(1-\frac{E^2-b^2}{b^2}\tan^2\varphi \right)} \right]\label{25},
\end{equation}
for $0\leq \varphi< \pi/2$. The above equation gives the relationship between the directions of the phase velocity and its associated group velocity.
The ``$\pm$'' signs mean that in general for a group velocity direction there are two associated wave vector directions. The converse is also true, as we commented before. For the case where Eq.~(\ref{17b}) is valid, we already know that the phase velocity cannot
be imaginary by any angle $\theta$ and it implies from Eq.~(\ref{25}) that
\begin{equation}
-\arctan\left(\frac{b}{\sqrt{E^2-b^2}}\right)< \varphi< \arctan\left(\frac{b}{\sqrt{E^2-b^2}}\right) \label{26},
\end{equation}
since ${\cal J}\geq 0$ by definition. Therefore, the group velocities are always restricted to a region of the $xz$ plane.
Fig.~\ref{fig2} depicts the above discussion for the same set of the parameters used in Fig.~\ref{fig1}. Birefringence is present for any chosen angle
inside the two thick straight lines. In other words, for any angle inside this region, two extraordinary rays (``$+$'' and ``$-$'') propagate in a same direction.
Outside this region, rays do not propagate.

\section{Discussion}
\label{discussion}

Phase and group velocities generally behave differently whenever the former is dependent upon the wave vector.
In this case, it is easy to show that the group velocity can be cast as
\begin{equation}
\vec{u}=v_{\phi}\hat q + \hat q \times \left(\frac{\partial v_{\phi}}{\partial \hat q}\times \hat q\right),
\end{equation}
where $v_{\phi}$ stands for any of the extraordinary phase velocities solutions to the Fresnel equation. Hence, whenever $v_{\phi}=v_{\phi}(\hat q)$,
an orthogonal term to the phase velocity appears in the group velocity. This is precisely what leads in general to a difference in the aforesaid velocities. Naturally this is the case in our analysis, as explicitly given by Eqs.~(\ref{21}) and (\ref{16}) and depicted in Figs.~\ref{fig1} and \ref{fig2} for a particular choice of the strength fields and $\alpha$.

Whenever one considers $E\leq b$, Eq.~(\ref{17b}) is trivially satisfied and one just has one refraction in terms of both waves and rays.
The associated extraordinary waves and rays tend to $\sqrt{\alpha}$ when $b$ goes to infinity, as expected. Nonetheless, setting $E>b$ and imposing real dielectric coefficients, a much richer scenario arises. In this configuration, there will always be a region where the one-way propagation phenomenon takes
place in terms of rays, accompanied by birefringence.

As we stressed previously, under certain conditions, there exists a correspondence between the Maxwell theory in a nonlinear material medium and a nonlinear theory of electromagnetism.
This theory, characterized by a given nonlinear Lagrangian $L$, emerges by the identifications \cite{BILag}
\begin{equation}
D_i=\frac{\partial L}{\partial E_i}= \sum_{j=1}^3\varepsilon_{ij}E_j,\;\; H_i=-\frac{\partial L}{\partial B_i}= \sum_{j=1}^3\mu^{\scriptscriptstyle{ -1}}_{ij}B_j.\label{dihi}
\end{equation}
In our case, the dielectric coefficients given by Eqs.~(\ref{03}) and (\ref{04}) do not in general admit a Lagrangian. This is so since the general condition fo guaranteeing the existence of $L$ in Eq.~(\ref{dihi}), viz.,
\begin{equation}
\frac{\partial D_i}{\partial B_j}=-\frac{\partial H_j}{\partial E_i},
\end{equation}
is not satisfied solely due to the fact that $\mu\epsilon \neq 1$ in general.
Nevertheless, for the particular case $\alpha=1\doteq 1/(\mu_0\varepsilon_0)$ (vacuum), we do have
\begin{equation}
L=\frac{b^2}{\mu_0}\left(1-\sqrt{1+\frac{F}{2\,b^2}-\frac{G^2}{16\,b^4}}\right)\label{lbi},
\end{equation}
where $F=2(|\vec{B}|^2-|\vec{E}|^2)$ and $G^2=16(\vec{E}\cdot\vec{B})^2$ are the two local invariants of the electromagnetism, and now the aforementioned medium parameter $b$ plays the role of the fundamental scale field to the theory. The above Lagrangian is analogous to the Born-Infeld Lagrangian \cite{BILag} and naturally raises from our analysis as a limiting case. It is worth mentioning that the Born-Infeld theory has recently been applied to the hydrogen atom \cite{carley2006,franklin2011}. It was concluded that both in the nonrelativistic and relativistic theories of quantum mechanics, the fundamental scale field to the Born-Infeld theory must be much larger than $ 10^{15} esu\simeq 10^{19}V/m$, which was determined by Born and Infeld themselves by assuming the unitarian viewpoint \cite{BILag}. Notwithstanding, a definite value for such a scale field was not found. The aforementioned viewpoint does not influence the derivation of the Born-Infeld theory and it has been assumed under more  philosophical grounds. Since quantum mechanics is basically founded on the dualistic viewpoint \cite{BILag} and due to the success of the former theory relying on the Maxwell Lagrangian for the electromagnetic fields, the above mentioned result is not surprising. Under the experimental point of view, such a fact makes the tests of the Born-Infeld Lagrangian even subtler, since fields as large as the ones mentioned above are at the present time unrealistic in terrestrial laboratories.

Notice that the denominators of Eqs.~(\ref{19}) and (\ref{20}) are proportional to the square-root term in Eq.~(\ref{16}). It means that whenever ${\cal J}$ is negative, the group velocities associated with Eqs.~(\ref{03}) and (\ref{04}) become superluminal,
whilst this is not the case for the phase velocities. If the dielectric tensors are real (losses are negligible and geometric optics is meaningful \cite{LandauCM1984}), it is then guaranteed that ${\cal J}$ is positive [see Eqs.~(\ref{24}) and (\ref{17b})] and hence superluminal group solutions do not rise in our model. Besides, our reasoning also implies that whenever Eq.~(\ref{17b}) is set, the controllable fields are not independent in the sense that the electric field must be dependent upon typically magnetic parameters, such as currents, and magnetic fields must depend upon charge densities. Indeed this is the case, since the associated field equations are nonlinear.

It can be shown  \cite{birula1970, delorenci2013} that the dielectric coefficients given by Eq.~(\ref{03}) and (\ref{04}) for $\alpha=1$ have a notable property: their associated Fresnel equation is independent of $G$. Therefore, the aforementioned media that have dielectric tensors with $\alpha\rightarrow 1$, but without the $G$ dependence, are also expected to display birefringence and one-way propagation effects. This could possibly be of experimental importance. Other powers of $G$ in Eqs.~(\ref{03}) and (\ref{04}) for $\alpha$ close to unity would lead to similar optical effects as the ones in the media sketched out before just in the vicinities of $G=0$.

Kruglov \cite{kruglov2010} has analyzed some wave aspects
of a modified version of the Born-Infeld Lagrangian, where there are two scale fields. When the scale fields are the same in his description,
birefringence disappears due to the simple fact it was assumed that the waves propagate just in an external magnetic field. This result
can be immediately seen from our description when $E=0$ [see Eq.~(\ref{16})].

If the media characterized by Eqs.~(\ref{03}) and (\ref{04}) could be tailored and fields inside them could be controlled by convenient charge densities,
then birefringence and one-way propagation of light are supposed to take place. Besides, as a quick glance in Eqs.~(\ref{03}), (\ref{04}), (\ref{17b}) and (\ref{26}) reveal, the underlying media must present some negative dielectric components for allowing the above-mentioned optical effects. This suggests that birefringence and one-way propagation of light could be found just in the realm of metamaterials and probably have not
been observed due to the very fine tuning of fields and tailored media they require. As a by-product of the aforementioned investigation, using continuity arguments for $\alpha$, one could indirectly assess the Born-Infeld Lagrangian in the realm of fields larger than its scale parameter. Direct investigations of such a Lagrangian would be possible just in the limiting case $\alpha$ tends to one, the case of a medium whose dielectric properties in the absence of controllable fields are close to the vacuum.

Anisotropic media could emerge from the so-called layered media \cite{pendry2006}. By a convenient choice of the base layers, nonlinear media that exhibit negative
dielectric coefficients can always be tailored. We hope this could be the case for the media proposed in this work, or they could rise by other means, as the technology of manipulating metamaterials is developing quickly \cite{smolyaninov2010}. Issues connected with losses in metamaterials are also of importance, since in the limit of geometric optics, wave and ray propagation are just meaningful in lossless media \cite{LandauEM1987}. It is known that significant progresses are being made in this direction \cite{boltasseva2011}. Applications concerning the optical effects analyzed in this work could be envisaged, for instance as optical diodes, due to the controlled unidirectional nature these metamaterials are expected to display either in terms of waves or rays. Besides, one could also in principle investigate analogously phenomena in black holes physics, as one-way propagation is also supposed to take place in such a scenario.

\acknowledgments
This work was partially supported by the Brazilian research agencies
CNPq, FAPEMIG and CAPES (under scholarship BEX 18011/12-8). J.P.P. acknowledges the support given by the
Erasmus Mundus Joint Doctorate Program, under the Grant No. 2011-1640
from EACEA of the European Commission.

\end{document}